\documentclass[10pt]{iopart}
\usepackage{epsfig}
\usepackage{graphicx}
\usepackage{dcolumn}
\usepackage{iopams}
\usepackage{color}

\expandafter\let\csname equation*\endcsname\relax 
\expandafter\let\csname endequation*\endcsname\relax 

\usepackage{amsmath}
\usepackage[english]{babel}
\usepackage[utf8]{inputenc}
\usepackage{hyperref}
\usepackage{bm}

\begin{document}

\title{Spin-Orbit Coupling in Deformed Harmonic Traps}

\author{O V Marchukov, A G Volosniev, D V Fedorov, A S Jensen and N T Zinner}
\address{Department of Physics and Astronomy, Aarhus University, 
DK-8000 Aarhus C, Denmark} 
\begin{abstract}
We consider a two-dimensional system of harmonically trapped particles with pseudo-spin-$\frac{1}{2}$ degree of freedom. This degree of freedom is coupled to the particle's momentum via the so-called Rashba spin-orbit interaction. We present our numerical results for a single-particle and few-particle systems, assuming the repulsive interparticle interaction to be of zero range. 
\end{abstract}
\maketitle
\section{Introduction}
\label{par:intro}
For the past several years the idea of quantum simulators has found its implementation in the ultracold atomic gas systems~\cite{bloch2012}. Trapped by an external magnetic or/and electric fields -- for example in an optical lattice -- ultracold gases are systems of extreme purity and tunability. Manipulation of external fields provides a possibility to simulate various phenomena from different areas of physics. Currently a lot of attention is attracted to the so-called synthetic gauge fields, see for example reviews~\cite{goldman2013, gerbier2013} and references therein, and in particular to the simulation of the spin-orbit coupling~\cite{lin2013}, which is -- in a broader sense -- a coupling between a particle's pseudo-spin and its motion. In our research we consider spin-orbit coupled particles in a trap and how the deformation of a trapping potential affects the energy spectrum. One can find an example of such a system in semiconductor nanostructure physics~\cite{avetisyan2012}.

\section{System.}
\label{par:system}
We consider a system of identical fermionic particles with a pseudo-spin-$\frac{1}{2}$ degree of freedom. The system is confined in a two-dimensional (2D) harmonic oscillator trap. We take the spin-orbit interaction in the so-called Rashba form~\cite{rashba1984}. We approximate the two-particle repulsive interaction via the contact potential, with the restriction that the particles can interact only in the spin singlet channel due to the Pauli exclusion principle. The Hamiltonian operator for such a system is written as
\begin{equation}
\label{eq:hamiltonian}
\hat H = \sum_{i=1}^{N} \left ( \frac{\mathbf {p_i^2}}{2m} + \frac{1}{2} m (\omega_x^2 x_i^2 + \omega_y^2 y_i^2) \right ) \otimes \hat I + \alpha_R \sum_{i=1}^{N} ({\hat \sigma_{xi}} {p_{yi}} - {\hat \sigma_{yi}} {p_{xi}}) + \frac{1}{2} g \sum_{i \neq j} \delta(\mathbf r_i - \mathbf r_j) \hat P_s,
\end{equation}
where $N$ is the number of particles, $\mathbf{p_i} = \{p_{xi}, p_{yi}\}$ and $\mathbf{r_i} = \{x_{i}, y_{i}\}$ are the 2D momentum and coordinate operators, $m$ is the particles' mass, $\omega_x$ and $\omega_y$ are the frequencies of the harmonic oscillator trap in the directions $x$ and $y$, $\hat I$ is the $2 \times 2$  unit matrix, $\alpha_R$ is the strength of the spin-orbit coupling, $\hat \sigma_x$ and $\hat \sigma_y$ are the $2 \times 2$ Pauli matrices, $g > 0$ is the strength of the interaction, $\delta(\mathbf r_i - \mathbf r_j)$ is the Dirac delta function and $\hat P_s$ is the spin singlet operator, which acts on spinor wave functions in the following way: $\hat P_s \bigl(\begin{smallmatrix}
a\\ b
\end{smallmatrix} \bigr)_i \bigl(\begin{smallmatrix}
c\\ d
\end{smallmatrix} \bigr)_j = \frac{1}{2} (a d - b c) \left [ \bigl(\begin{smallmatrix}
1\\ 0
\end{smallmatrix} \bigr)_i \bigl(\begin{smallmatrix}
0\\ 1
\end{smallmatrix} \bigr)_j - \bigl(\begin{smallmatrix}
0\\ 1
\end{smallmatrix} \bigr)_i \bigl(\begin{smallmatrix}
1\\ 0
\end{smallmatrix} \bigr)_j  \right ]$, where indices $i$ and $j$ ($i \neq j$) show that the spinors correspond to different particles.

\section{Energy spectrum.}
\label{par:spectrum}
The single-particle case was previously investigated by means of an exact diagonalization method~\cite{marchukov2013}. To investigate the system of $N$ fermions we implement the Hartree-Fock mean field method. We consider an ansatz that the $N$-body wave function is a Slater determinant -- an antisymmetrized product of the single-particle wave functions. In order to find the approximation to the ground state energy of the system we minimize the energy functional $E[\Psi] = \frac{\langle \Psi \mid \hat H \mid \Psi \rangle}{\langle \Psi \mid \Psi \rangle}$, where $\Psi$ is the Hartree-Fock $N$-body wave function and $\hat H$ is the system's Hamiltonian. This minimization gives the self-consistent Hartree-Fock equations, that can be solved iteratively: on every step we, using the same routines that we used for the single-particle case, obtain a set of eigenstates and eigenvalues which are used to construct the $N$-body wave function for the next iteration. The procedure continues until the ground state energy is converged. Fig. 1 shows the energy levels of the system as function of the dimensionless spin-orbit strength parameter, $\alpha_R / \sqrt{\frac{m}{2\hbar \omega_y}}$, for different deformations. One can see that the density of states of both single-particle and N-particle systems depend on the deformation of the trap. For a strong deformation the lowest levels are equidistant and show the one-dimensional behavior -- the energy decreases as $- m \alpha_R^2 / 2$. In the case when the ratio of the frequencies is rational but not an integer number (center panel of Fig. 1) the states are more evenly distributed. We see that the Hartree-Fock single-particle energy levels are shifted compared to the single-particle case, but the overall structure stays very similar. This spectrum is a special case of the so-called Fock-Darwin spectrum~\cite{reimann2002}.
\begin{figure*}
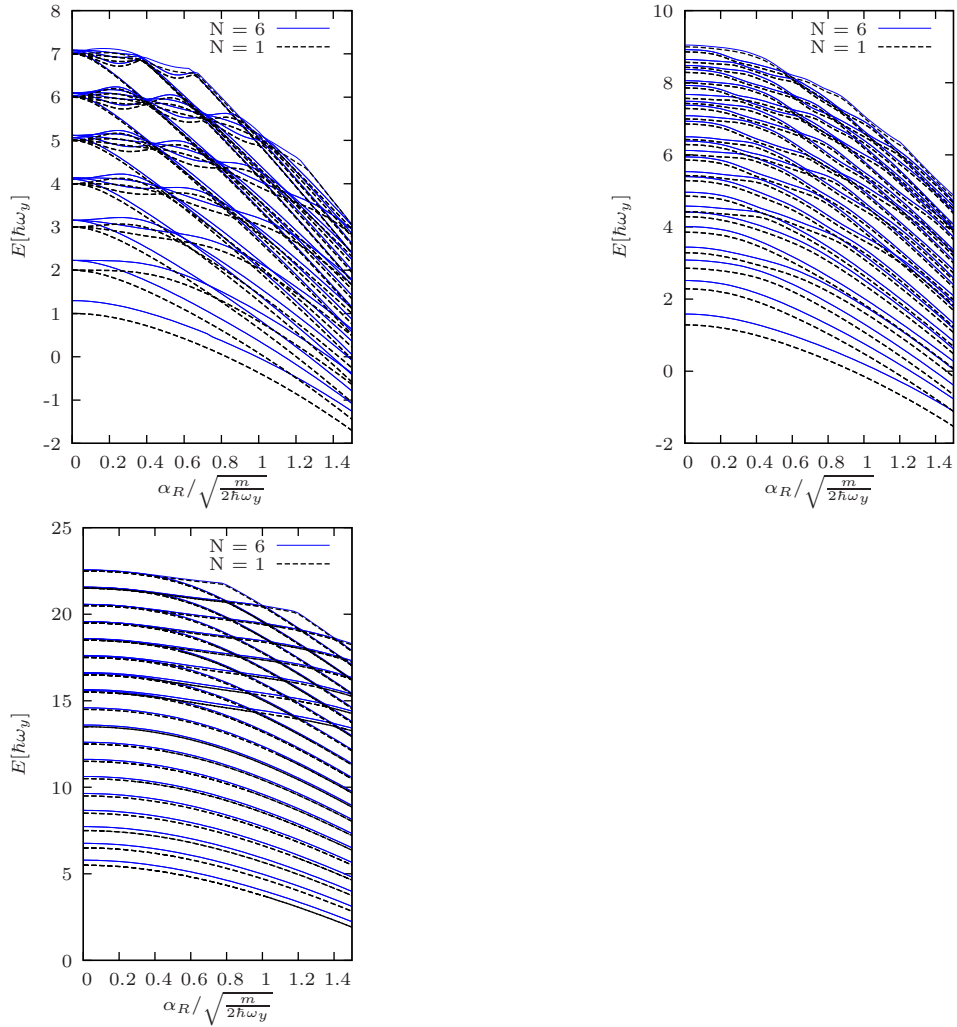

\label{fig:sp} 
  \input{fig1a.tex}
  \input{fig1b.tex}
  \input{fig1c.tex}
\caption{The first 50 energy levels for a single-particle case (black dashed) and the Hartree-Fock single-particle energy levels (blue solid) as a function of the dimensionless spin-orbit coupling strength, $\alpha_R / \sqrt{\frac{m}{2 \hbar \omega_y}}$ . The interaction strength $g = \frac{\hbar^2}{m}$. The deformations of the harmonic trap are different: rotationally symmetric $\omega_x = \omega_y$ (left), small rational deformation $\frac{\omega_x}{\omega_y} = 1.57$ (center) and large deformation $\frac{\omega_x}{\omega_y} = 10$ (right).}
\end{figure*}

\end{document}